\documentclass[pra,twocolumn,showpacs]{revtex4}
\usepackage{subfigure}
\usepackage{graphicx}

\begin{document}
\title{Single Cs Atoms as Collisional Probes in a large Rb Magneto-Optical Trap}
 \author{Claudia Weber}
    \author{Shincy John}
    \author{Nicolas Spethmann}
    \author{Dieter Meschede}
    \author{Artur Widera} \email{widera@uni-bonn.de}
    \address{Institut f\"ur Angewandte Physik, Universit\"at Bonn,
    Wegelerstr.~8, D-53115 Bonn, Germany}

\begin{abstract}
We study cold inter-species collisions of Caesium and Rubidium in
a strongly imbalanced system with single and few Cs atoms.
Observation of the single atom fluorescence dynamics yields
insight into light-induced loss mechanisms, while both subsystems
can remain in steady-state. This significantly simplifies the
analysis of the dynamics, as Cs-Cs collisions are effectively
absent and the majority component remains unaffected, allowing us
to extract a precise value of the Rb-Cs collision parameter.
Extending our results to ground state collisions would allow to
use single neutral atoms as coherent probes for larger quantum
systems.
\end{abstract}
 \maketitle
\section{Introduction}
Magneto-optical traps (MOTs) are standard sources for  ultracold
atoms \cite{Weiner_RMP71_1999}, and used with multiple species in
current experiments \cite{Taglieber2006,
spiegelhalder_all-optical_2010} as a first step toward the
creation of, e.g., ultracold hetero-nuclear molecules
\cite{kerman_production_2004, kraft_formation_2006, Voigt2009}. In
these MOTs, inter-species collisions can lead to dramatic losses
compared to the homo-nuclear case, limiting the density of atoms
available and complicating the experimental procedure. For various
single-species MOTs the homo-nuclear loss mechanisms are well
known as well as the corresponding interaction potentials.
However, this knowledge cannot be easily transferred to the
hetero-nuclear case: While the loss mechanisms remain the same
ones, their contributions to the dynamics of the mixed species
system can be strongly altered as the inter-nuclear long-range
potential is weaker and short range interaction dominates. The
loss mechanisms include ground state collisions due to spin
exchange and dipolar relaxation interaction as well as
light-induced collisions (ground-excited collisions)
\cite{Weiner_RMP71_1999}. The latter processes such as radiative
escape and fine-structure changing collisions dominate in traps
with near resonant light \cite{Gensemer_PRA62_2000,
Gallagher_PRL63_1989, Sesko_PRL63_1989}.
\par
The standard methods to extract information about these
inter-species interaction processes either consider the loading
dynamics of one atomic species in the MOT upon presence of
another, or investigate the change in trapped atom number as one
species is removed from the system which was in steady-state
previously \cite{Mancini_EPJD30_2004}. In both cases, the
simultaneous presence of homo-nuclear and hetero-nuclear
collisions, both depending on the atomic densities of the
respective species, complicate the analysis. In some cases, the
losses due to homo-nuclear collisions are neglected if the
corresponding loss rate $\beta_{i(j)}$ is much smaller than the
inter-species loss rate $\beta^\prime_{i,j}$ \cite{Footnote3}.
\par
Here we consider ground-excited collisions in a mixture of Caesium
(Cs) and Rubidium (Rb) \cite{Telles2001, Holmes2004} using a
system which combines the advantages of both methods mentioned
above. We deduce the loading and loss dynamics of single Cs atoms
in the presence of a large Rb MOT, while both sub-systems are in
steady state. Working with single atoms, the probability for
homo-nuclear Cs-Cs collisions is negligible simplifying the
analysis. In principle, however, our method even allows
identifying and counting these rare events. At the same time, the
presence of single Cs atoms has no measurable effect on the Rb
system.
\begin{figure}\centering{
\includegraphics[width=1\linewidth]{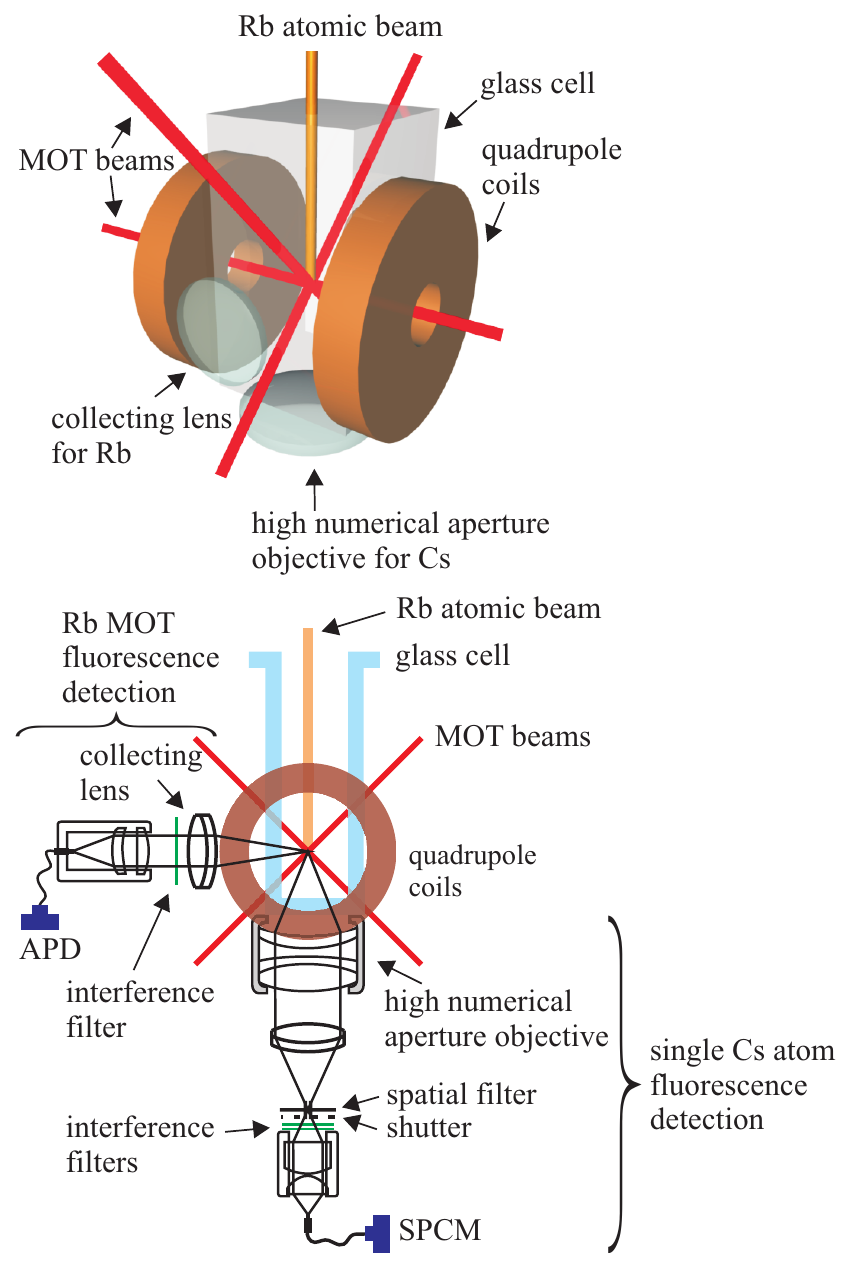}}
\caption{(Color online) Schematic of the experimental setup to
operate and detect a strongly imbalanced two-species MOT. Both
MOTs of Rb and Cs are produced at the same position using the same
quadrupole coils. The Rb MOT is loaded by an atomic beam (orange)
provided by a second vapour pressure MOT, whereas Cs is loaded
from the background of the ultra-high vacuum. The Rb and Cs atom
numbers are determined by fluorescence detection setups located at
the side and at the bottom of the glass cell, respectively. The Cs
fluorescence light collected by a high numerical aperture
objective is detected by an APD-SPCM with a quantum efficiency of
42\%. To suppress stray light as well as Rb fluorescence and laser
light a 300\,$\mu$m pinhole and two interference filters serve as
spatial and spectral filters, respectively. The Rb fluorescence is
collected by a lens, spectrally filtered and detected by an
APD.\label{Skizze}}
\end{figure}

\section{Experimental setup}
In order to probe cold collisions of the two-species mixtures with
a single atom probe, we trap Rb and Cs simultaneously but in
different regimes of atom numbers.
\par
\begin{figure*}\centering
\includegraphics[width=1\linewidth]{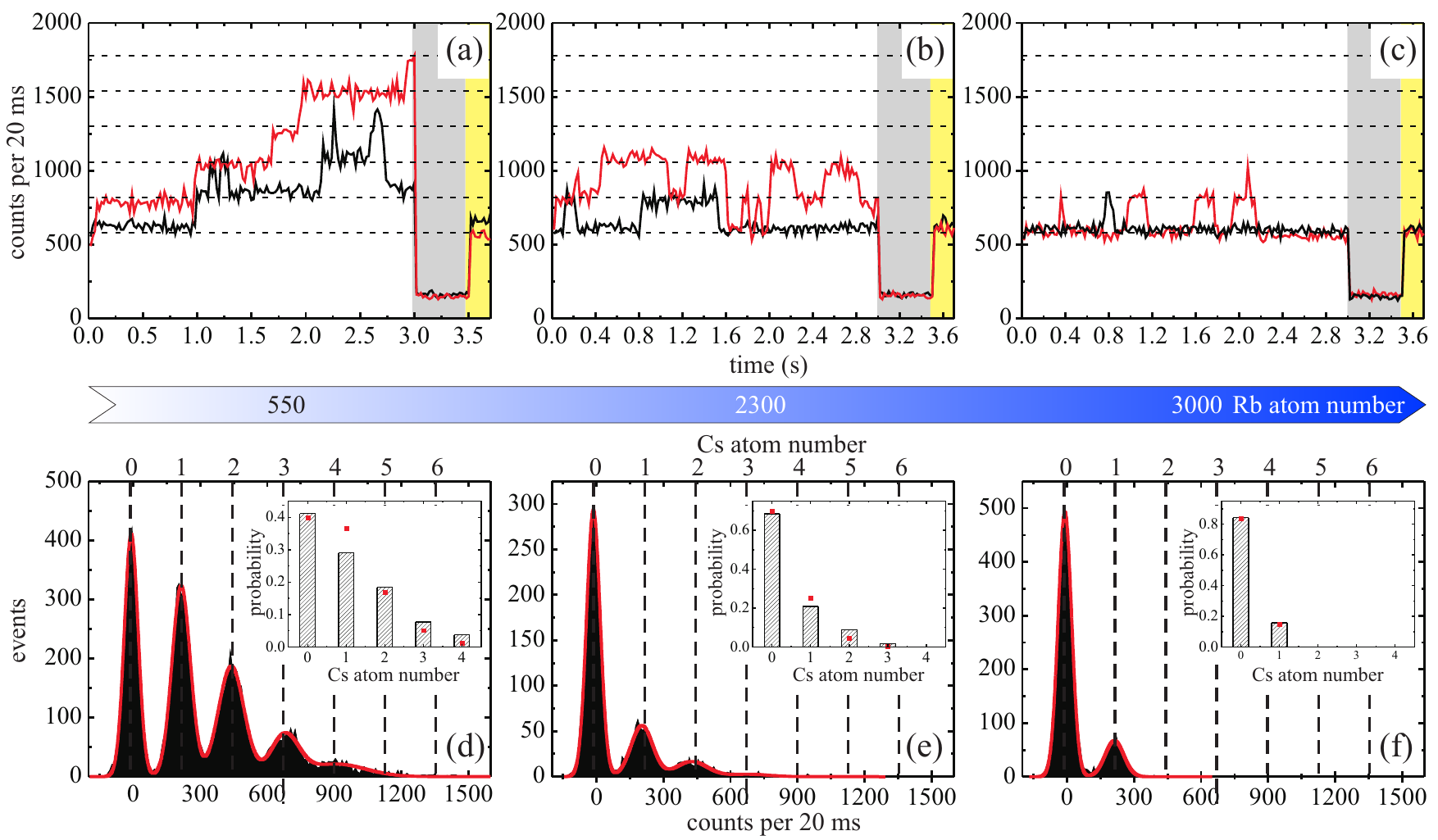}\caption{(Color online) (a)-(c) Typical fluorescence traces of the single Cs atoms fluorescence signal for a
selection of three different Rb atom numbers: The detected pulses
are binned in time intervals of 20\,ms each.  Each panel shows two
typical fluorescence traces of single Cs atoms in the presence of
about 550, 2300 and 3000\,Rb atoms, respectively. Within the grey
region the cooling laser light and the magnetic field is switched
off. For the last 200\,ms (yellow shading) the laser light is
switched on again to determine the background light level. The
Cs-Cs collision event shown in figure (a) at 2.5\,s is one of the
very few observed ones. (d)-(f) Corresponding histograms of the Cs
fluorescence signal for the same Rb atom number as in (a)-(c).
Each peak corresponds to 0,1,2,... atoms, respectively. The red
lines are Gaussian fits to the peaks. We find that, not only in
steady-state, but for all data sets the probability distribution
can be well described by a Poissonian function. The insets compare
the measured probability distribution (bars) with a fitted
Poissonian distribution (red dots). \label{histotrace}}
\end{figure*}

We load single Cs atoms from a low background pressure well below
$10^{-11}$\,mbar in a high magnetic field gradient magneto-optical
trap (MOT) \cite{Monroe_PRL65_1990, Haubrich_EPL34_1996}. The
relatively high magnetic field gradient of 270\,G/cm reduces the
loading rate compared to a standard MOT by several orders of
magnitude to about 1~atom/s. In order to support a low loading
rate the MOT is operated using relatively small MOT beams with a
diameter of 1\,mm and a total power of 500\,$\mu$W. The atom
number in the MOT is detected by a sensitive fluorescence
detection system schematically drawn in Fig.~\ref{Skizze}. It is
based on a high numerical aperture objective (NA=0.29), spectral
and spatial filters and an avalanche photodiode operating in a
single photon counting mode (APD-SPCM). In this operating mode,
also known as Geiger mode, each detected photon produces a TTL
pulse which is accumulated by a counter card. The field-of-view of
the detection setup is about 60\,$\mu$m in the object plane, which
is approximately twice the maximum of the diameter of the high
gradient Cs MOT. We record the fluorescence for up to 4\,s, where
the photon count rate is binned in time intervals of 20\,ms. The
detection time is limited by the power dissipation of the
quadrupole coils. Then, all cooling laser beams and the magnetic
field are switched off for 500\,ms so that all atoms escape from
the trap. Subsequently, the MOT laser beams are switched on
without magnetic field applied, yielding the background light
level due to stray light. Typical fluorescence traces are shown in
Fig.~\ref{histotrace}. The discrete fluorescence steps correspond
to the loading or loss of one atom \cite{Ruschewitz_EPL34_1996,
Haubrich_EPL34_1996}. The photon count rate per atom is
experimentally determined to be $10^4$\,counts s$^{-1}$. A
corresponding histogram of 200 such traces showing the occurrences
of each count rate for the same set of experimental parameters is
shown in Fig.~\ref{histotrace}, where the background rate has been
subtracted individually for each trace to eliminate long time
drifts. Each peak is attributed to a number of atoms $N$. In
steady-state the probability distribution of the detected atom
number is described by a Poissonian distribution.
\par
The Rb MOT is superposed to the single Cs MOT by using the same
magnetic quadrupole field. Both MOTs can have strongly different
atom numbers (up to $3\times 10^3$ atoms for Rb) due to different
parameters of the corresponding laser beams. The large wave length
difference of 72\,nm between the atomic transitions for Rb
(780\,nm) and Cs (852\,nm) allows adjusting the MOT laser system
for each species without affecting the other. For Rb we choose a
beam diameter of 16\,mm with a power of 100\,mW. In addition, the
Rb MOT is loaded from an atomic beam originating from a second
vapour pressure MOT (see Fig. \ref{Skizze}). This atomic beam
provides precooled Rb atoms and leads to a local enhancement of
the partial Rb background pressure at the MOT position. Thereby
the loading rate increases by up to three orders of magnitude
compared to the single atom MOT, resulting in a higher number of
stored Rb atoms of up to 3300 atoms. This corresponds to a peak
density of $6.5\cdot10^{10}$\,cm$^{-3}$ which is comparable to
standard MOT densities. During the experimental sequence an
adjustment of the number of Rb atoms trapped is performed by
changing the atomic flux of this atom beam without affecting the
MOT properties such as position or size. As in the case of Cs the
number of Rb atoms is determined by a fluorescence detection
system. Due to the much larger atom number of Rb compared to Cs,
the corresponding detected fluorescence rate is several orders of
magnitude larger for Rb. Therefore, the avalanche photodiode (APD)
is operated below the Geiger mode \cite{Footnote1}, and thus the
photon number is directly proportional to the output current, and
we obtain direct information about the fluorescence intensity. The
absolute number of the Rb atoms is only precise to a factor of
1.3. As schematically drawn in Fig.~\ref{Skizze} the determination
of the Rb atom number depends on the effective solid angle of
detection, on the reflectivity and absorption of the optics used,
the fibre coupling efficiency and the responsivity of the APD. The
main influence are the two last points due to the non-linear
dependence of the APD responsivity on the operating voltage
$(40\pm5)$\,A/W and the estimated fibre coupling efficiency of
$(60\pm20)\,\%$.
\par
In order to observe interaction between Rb and Cs, we first load
Rb atoms in the high gradient magnetic field MOT for 2.5\,s until
the Rb atom number is in steady-state for a chosen number of Rb
atoms. Once this is reached, the Cs MOT-light is switched on and
the fluorescence detection signal of Cs is recorded for 3\,s. As
mentioned above, we then detect the dark count rate and background
light level for each trace.
\section{Analysis of Cs atom dynamics}
Cs fluorescence traces have been recorded for Rb atom numbers in
the range between 0 and 3300\,Rb atoms. All data are sorted by
$N_{\rm{Rb}}$, which is binned in steps of 220\,Rb atoms, and for
each Rb atom number we average over typically 200 traces
\cite{Footnote2}. An example of two typical traces for three
different Rb atom numbers is plotted in Fig.~\ref{histotrace}.
Qualitatively different dynamics of the single Cs atom MOT can
already be seen in the single recorded traces. For an almost
negligible Rb atom number (Fig.~\ref{histotrace}\,a) a fast
loading of Cs can be observed with only few loss features which
does not saturate during our observation time. In contrast, only a
few loading events can be found for large Rb atom numbers
(Fig.~\ref{histotrace}\,c). Moreover, loss processes are much more
frequent and dominated by single-Cs collisions. All the loading
events are followed, within 100\,ms, by an atom loss. Within this
range of the Rb atom number many
traces do not show any event at all.\\
The corresponding histograms shown in Fig.~\ref{histotrace}\,d-f
mirror this behavior. In Fig.~\ref{histotrace}\,d for about
$N_{\rm{Rb}}=550$ five peaks corresponding to up to four atoms can
be observed, whereas in Fig.~\ref{histotrace}\,f at most one atom
is detected \cite{Footnote3}.
\\
In order to get information about the inter-species collision
properties the dynamics of the Cs-MOT is analyzed. In the presence
of Rb this dynamics can be described by the rate equation
\begin{eqnarray} \label{ratengl1}
\frac{dN_{\rm{Cs}}}{dt} &=& R(N_{\rm{Rb}})-\gamma\,
N_{\rm{Cs}} \nonumber \\
& & -\beta_{\rm{RbCs}}\int n_{\rm{Rb}}(r,t)\,
n_{\rm{Cs}}(r,t)\,d^{3}r  \nonumber \\
& & -\beta_{\rm{CsCs}}\int
n^2_{\rm{Cs}}(r,t)\,d^{3}r.
\end{eqnarray}

\begin{figure}\centering
\includegraphics[width=1\linewidth]{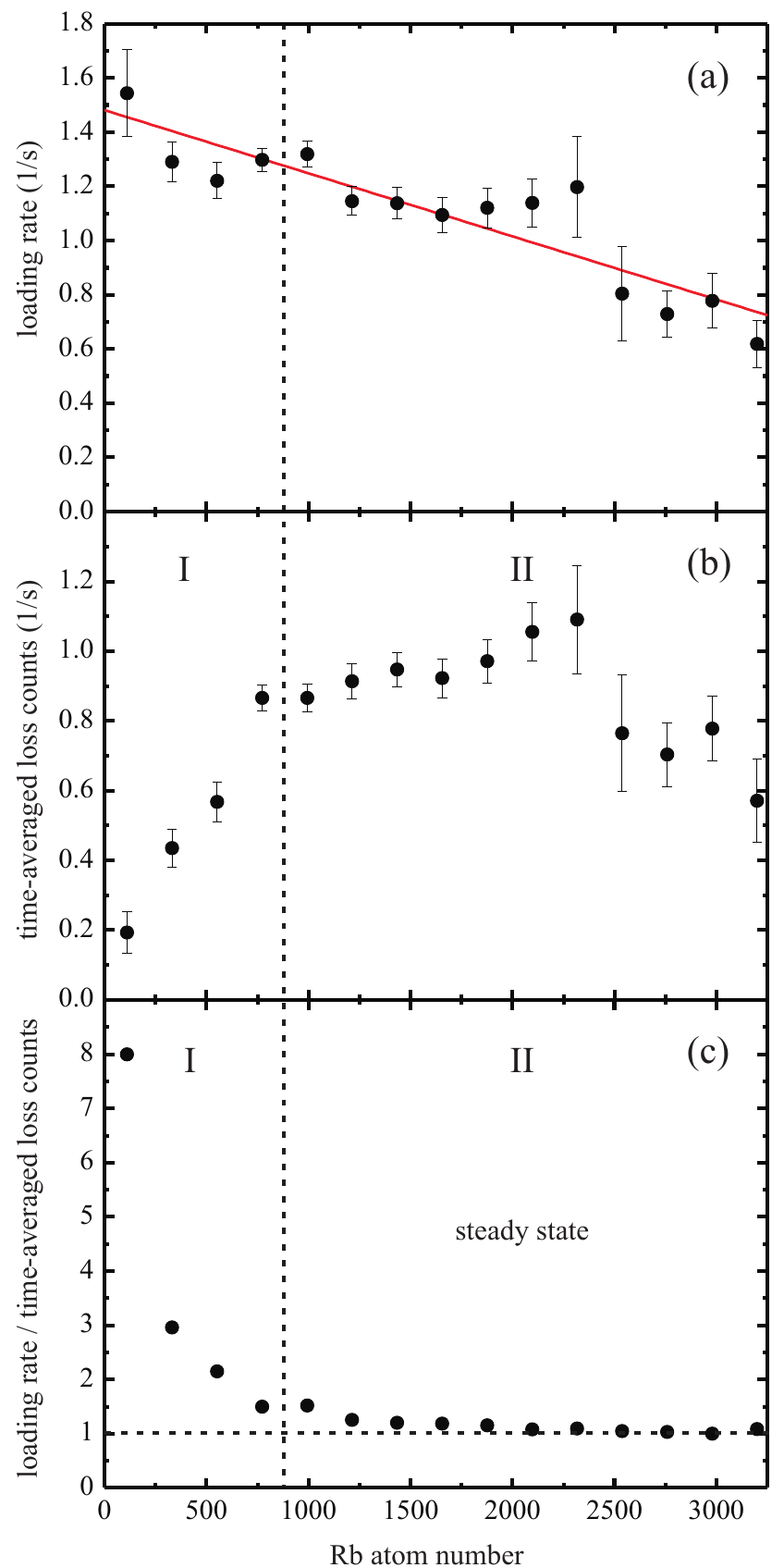}\caption{(Color online) Dynamics of the Cs MOT depending on the number of Rb atoms trapped: (a) Loading rate. The red line is a linear fit to the data
yielding $R_{0}=(1.48\pm0.06)$\,s$^{-1}$ and $\alpha=(2.3\pm
0.3)\cdot10^{-4}$\,s$^{-1}$. (b) Time-averaged loss counts, and
(c) ratio of loading rates and time-averaged loss counts. The
vertical dashed line divides the graphs into two sections. Within
section I the loading rate dominates the Cs MOT dynamics. For
larger Rb atom number (section II) the ratio between loading rates
and time-averaged loss counts tends to unity (horizontal dashed
line), which corresponds to Cs being in steady
state.\label{ladeverlust}}
\end{figure}
Here, the first term is the Rb atom number dependent loading rate
$R(N_{\rm{Rb}})$; the second term describes the loss of Cs atoms
due to collisions with background gas particles at a rate
$\gamma$; the third and fourth terms describe, respectively, the
loss of Cs atoms due to collisions with a Rb and Cs atom,
characterized by the inelastic collision coefficients
$\beta_{\rm{RbCs}}$ and $\beta_{\rm{CsCs}}$. In our particular
system of single Cs atoms, the different terms can be determined
either directly by the recorded fluorescence traces, by
independent measurements or can be neglected. In contrast to
common balanced many-body mixtures, here only the value of the
inelastic Rb-Cs collision coefficient $\beta_{\rm{RbCs}}$ remains
unknown. In the following each term is discussed separately.

The background collision rate $\gamma$ is determined by an
independent measurement in the absence of Rb to be $(0.03\pm
0.01)$\,s$^{-1}$. The last term describing Cs-Cs collisions which
rarely occur due to the extremely low Cs density and lead to a
simultaneous loss of two Cs atoms can be neglected. This is an
important advantage in comparison to balanced many-body mixtures,
where intra-species interactions have to be taken into account.
\begin{figure}\centering
\includegraphics[width=1\linewidth]{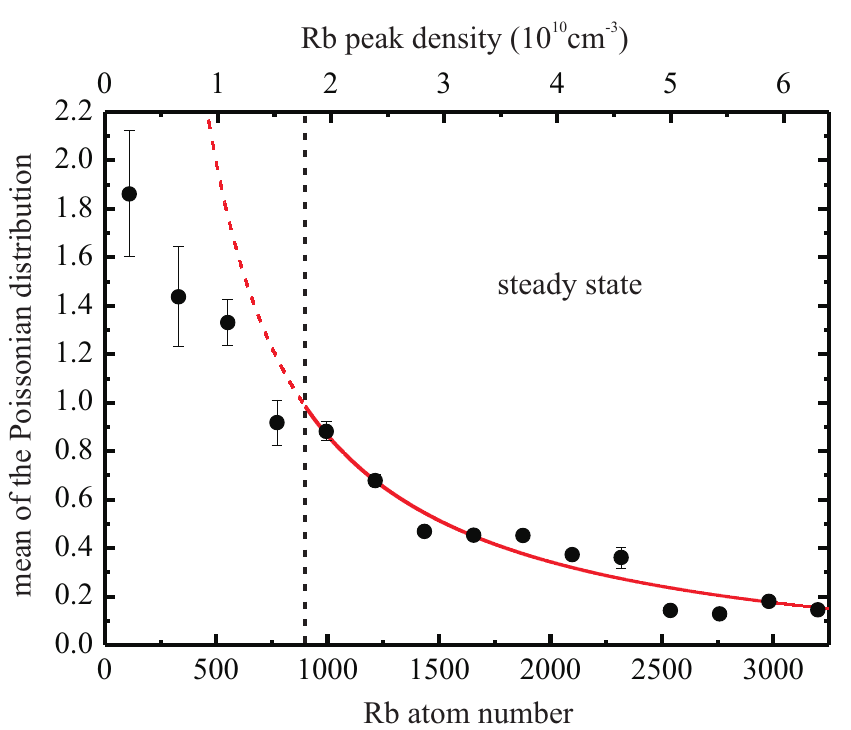}\caption{(Color online) Mean of the Poissonian distribution of Cs atom numbers as a
function of the Rb atom number and peak density. The red solid
line is a fit of equation (\ref{CSAtomanzahl}) to the data within
the steady state regime. The red dashed line is the extrapolation
to the non-steady state regime of the Cs MOT.
\label{ErwartungswertmitRb}}
\end{figure}
\par
In order to determine the first term in Eq.~\ref{ratengl1}, we
obtain the loading rate of the detected traces by counting the
loading steps and averaging over the detection time. This is
plotted as a function of the Rb atom number in
Fig.~\ref{ladeverlust}\,a, showing a linear decrease of the
loading rate for an increasing number of Rb atoms described by
$R(N_{\rm{Rb}}) =R_{0}-\alpha\,N_{\rm{Rb}}$, where $R_{0}$ denotes
the loading rate of Cs without Rb and $\alpha$ is the
proportionality factor. A linear fit determines the parameters to
be $R_{0}=(1.48\pm0.06)$\,s$^{-1}$ and $\alpha=(2.3\pm 0.3) \times
10^{-4}$\,s$^{-1}$. We attribute the decreasing loading rate with
increasing Rb atom number to the fact that the relatively large Rb
MOT operates as a shield for the Cs MOT. The enhanced Rb partial
background pressure surrounding the Cs MOT in combination with the
MOT laser beams leads to light-induced cold collisions during the
capturing process of Cs atoms thereby reducing the Cs loading rate
for an increasing number of Rb atoms stored.
\par
Analogous to the loading rate, the loss counts per time are
determined averaging over the detection time, which is in general
different from the loss rate, i.e.~the inverse lifetime of an atom
in the trap. The time-averaged loss counts, which are plotted in
Fig.~\ref{ladeverlust}\,b, however, do not show a monotonic
behaviour, which can be explained by regarding the ratio of the
loading rate to the time-averaged loss counts presented in
Fig.~\ref{ladeverlust}\,c.
Within the first region (I) for Rb atom numbers of up to 1000, the
loss counts per time increase with a rising number of Rb atoms.
Here, the loading rate dominates the dynamics of the single Cs
atom MOT. However, although the loading rate decreases, leading to
less captured Cs atoms, the loss counts per time is seen to
increase. Hence, the enhancement of losses has to be a two species
effect and can be attributed to cold collisions between Cs and Rb
discussed in \cite{Weiner_RMP71_1999}. In this range of Rb atom
numbers, where the loading rate exceeds the value of the
time-averaged loss counts, the measured mean Cs atom number
depends on the loading and detection time, which can also be seen
in the fluorescence traces of Fig.~\ref{histotrace}\,a. The Cs
atom number increases continuously over the detection time without
reaching a final steady-state value until the MOT is switched off.
\par
For higher Rb atom numbers, i.e.~in section II, the ratio between
loading rate and loss counts per time tends to unity as shown in
Fig.~\ref{ladeverlust}\,c. Therefore, the evolution of the
time-averaged loss counts is determined by the loading rate.
Within this section on average each loaded atom leaves the MOT
again during the detection time. In this regime the Cs system is
in steady state, implying that the average Cs atom number does not
change with an increasing loading and detection time.
\par
Thus, for a Rb atom number larger than 1000, the time derivative
of the mean Cs atom number $\bar{N}_{\rm{Cs}}$ can be set to zero
and the steady state number $\bar{N}_{\rm{Cs}}$ of Cs atoms is
given by a simplified version of equation (\ref{ratengl1}) as
\begin{eqnarray} \label{CSAtomanzahl}
\bar{N}_{\rm{Cs}} &=&\frac{R_{0}-\alpha\,
N_{\rm{Rb}}}{\gamma+\beta_{\rm{RbCs}}\,
\frac{N_{\rm{Rb}}}{(17\,\pi)^{3/2}\,w_{\rm{Cs}}^{3}}}.
\end{eqnarray}
Besides the steady state regime of Cs, here, also the Rb atom
number is assumed to be time-independent. In fact, we have not
observed any effect of the presence of single or few Cs atoms on
the overall state of the Rb cloud with several hundred Rb atoms.
However, Rb-Rb cold collisions and collisions with background
particles limit the lifetime of the Rb MOT, thereby reducing the
Rb atom number. Experimentally the total Rb atom number is
maintained by continuously reloading Rb atoms from the atomic
beam.
Therefore, in equation \ref{CSAtomanzahl} a time-independent
Gaussian density distribution of the two clouds is assumed
\cite{Weiner_RMP71_1999}
\begin{eqnarray} \label{density}
 n_{i}(r,t) &=& n_{i}(r) =n^{0}_{i}\cdot
 e^{-\frac{r^2}{w_{i}^{2}}},
\end{eqnarray}
where $w_i$ is defined as the 1/\emph{e}-radius of the cloud of
species $i$, and $n^0_i$ is the corresponding central density. The
mean radius of the Cs cloud is calculated to be
$w_{\rm{Cs}}=6.6\,\mu$m following \cite{Metcalf__1999} and
averaging about all Zeeman states. Here, the magnetic field
gradient as well as the laser beam parameters, such as intensity,
waist, and detuning are taken into account. For Rb the size of the
cloud is four times larger, mainly due to a larger MOT beam size
and a different cooling light detuning. The occupation of all
different $m_F$-states results in an uncertainty of 15\,\% of the
absolute value of the size of the clouds.
\\
The steady state Cs atom number $\bar{N}_{\rm{Cs}}$ corresponds to
the mean value of Poissonian distributed Cs atom numbers measured
for one set of experimental parameters. 
The resulting expectation values of Cs are plotted versus the
number of Rb atoms in Fig.~\ref{ErwartungswertmitRb}. A strong
decay of the expectation value with increasing $N_{\rm{Rb}}$ can
be observed.
We fit Eq.~(\ref{CSAtomanzahl}) to the data within the steady
state regime to deduce $\beta_{\rm{RbCs}}$ as the only free
parameter. Extrapolating the fit to the region below 1000\,Rb
atoms, where the Cs system is not in steady state, the data and
the fit deviate. Here, the equilibrium Cs atom number given by the
extrapolation exceeds the measured expectation value, as the
measured Cs atom number is limited by the detection time. The
inelastic inter-species collision coefficient is determined to be
$\beta_{\rm{RbCs}}=(1.6\pm 0.3)\cdot 10^{-10}$\,cm$^{3}$/s. The
error is given by the statistical uncertainty of the fit. It has
been shown \cite{Sesko_PRL63_1989} that the absolute value of the
collision coefficient depends on the laser beam parameters such as
intensity and detuning. In principle this can also be investigated
by our single atom method. A systematic deviation of the
determined collision coefficient is obtained to be
$9\cdot10^{-11}$\,cm$^{3}$/s. This value is dominated by the
uncertainties of the Rb atom number and the size of the clouds as
discussed above.
The inelastic interspecies collision coefficient
$\beta_{\rm{RbCs}}$ includes all mechanisms of inelastic cold
collisions summarized as ground state and light induced
collisions.
Our measured value is in good agreement with Rb-Cs collision rate
measurements done in a balanced Rb-Cs MOT in
\cite{Harris_JPB41_2008} where intra-species interactions can not
be neglected. Our work thereby confirms that a single atom is
sufficient to probe large many-body systems.

\section{Conclusion}
We have illustrated the use of single neutral atoms
as a sensitive probe to investigate many-body systems, where
the overall state of the many-body system remains unmodified. The
inelastic Rb-Cs cold collision coefficient has been extracted from
the dynamics of the single atom MOT. As a crucial advantage over
balanced systems only inter-species interactions have to be
considered thereby simplifying the analysis.
\par
This experiment is a step en route towards controlled doping of
BECs with impurity atoms as a probe to investigate, e.g., the
decoherence of BECs \cite{Ng_PRA78_2008} or its phase fluctuations
\cite{Bruderer_NJP8_2006}. Since for the presence of near resonant
light it has been shown that the atomic interaction is dominated
by light-induced collisions which involve higher electronic states
\cite{Gensemer_PRA62_2000, Gallagher_PRL63_1989,
Sesko_PRL63_1989}, a crucial step is to constrain the interaction
to coherent ground state collisions by storing both species in
far-off resonant dipole traps. For these species a significant
elastic inter-species interaction strength was found
\cite{Anderlini2005, Haas_NJP9_2007}. In order to investigate the
effect of ground state collisions, time periods involving photon
scattering, e.g.~preparation or detection, must be temporally and
spatially separated from time intervals of ground state
collisions. Here, Feshbach resonances can be exploited to tune the
Rb-Cs interaction strength \cite{Pilch_PRA79_2009} enabling
interactions in only preset periods of time.

\section*{Acknowledgments}
We thank T.~Weikum for contributions to the fluorescence detection
system, and W.~Alt, O.~Fetsch and A.~Moqanaki for helpful
discussions. We gratefully acknowledge financial support from the
science ministry of North Rhine-Westphalia (NRW MWIFT) through an
independent Junior Research Group. N.S.~acknowledges support from
Studienstiftung des deutschen Volkes and N.S.~and C.W.~from the
Bonn Cologne Graduate School.

\begin{thebibliography} {99}
\bibitem{Footnote3} {It has been
    pointed out that the hetero nuclear loss rates are not reciprocal,
    i.e.~one cannot a priori assume
    $\beta^\prime_{i,j}=\beta^\prime_{j,i}$.}
\bibitem{Footnote1} {The Geiger mode can only be used for photon rates which are smaller than the inverse
    dead time of 1/35\,ns of the APD. Considering the corresponding
    detection setups the count rates of Rb and Cs are $10^8$/s and
    $10^5$/s, respectively, thus only allowing to operate the Cs APD
    in the Geiger mode.}
\bibitem{Footnote2} {For Rb atom numbers less than 700 the statistics is
    much less, and we average over about 50 traces only. However, this
    does not influence the analysis of the Rb-Cs interaction parameter,
    because this is based only on data sets with Rb atom numbers larger
    than or equal to 1000.}
\bibitem{Footnote3} {While in steady state the histograms are expected to follow a
    Poissonian distribution, we find that also for non-steady state
    conditions our data are very well described by a Poissonian. For
    simplicity we therefore fit a Poissonian distribution to the data
    and extract its mean atom number which is shifted towards smaller
    values for an increasing number of Rb atoms.}
\end{thebibliography}
\end{document}